\documentclass{article}

\usepackage{arxiv}

\usepackage[utf8]{inputenc} 
\usepackage[T1]{fontenc}    
\usepackage{hyperref}       
\usepackage{url}            
\usepackage{booktabs}       
\usepackage{amsfonts}       
\usepackage{nicefrac}       
\usepackage{microtype}      
\usepackage{lipsum}
\usepackage{graphicx}
\usepackage{enumerate} 
\usepackage{paralist}
\usepackage{listings}
\usepackage{multirow}
\usepackage{subfig}
\usepackage{graphicx}
\usepackage{tikz}
\usepackage{url}
\usepackage{xcolor}
\usepackage{soul}
\usepackage{xargs}
\usepackage{balance}
\usepackage{multirow}
\usepackage{enumitem}
\usepackage{footmisc}
\usepackage{parcolumns}
\usepackage{float}
\graphicspath{ {./images/} }

\newcommand{\orcid}[1]{\href{https://orcid.org/#1}{\includegraphics[height=10pt]{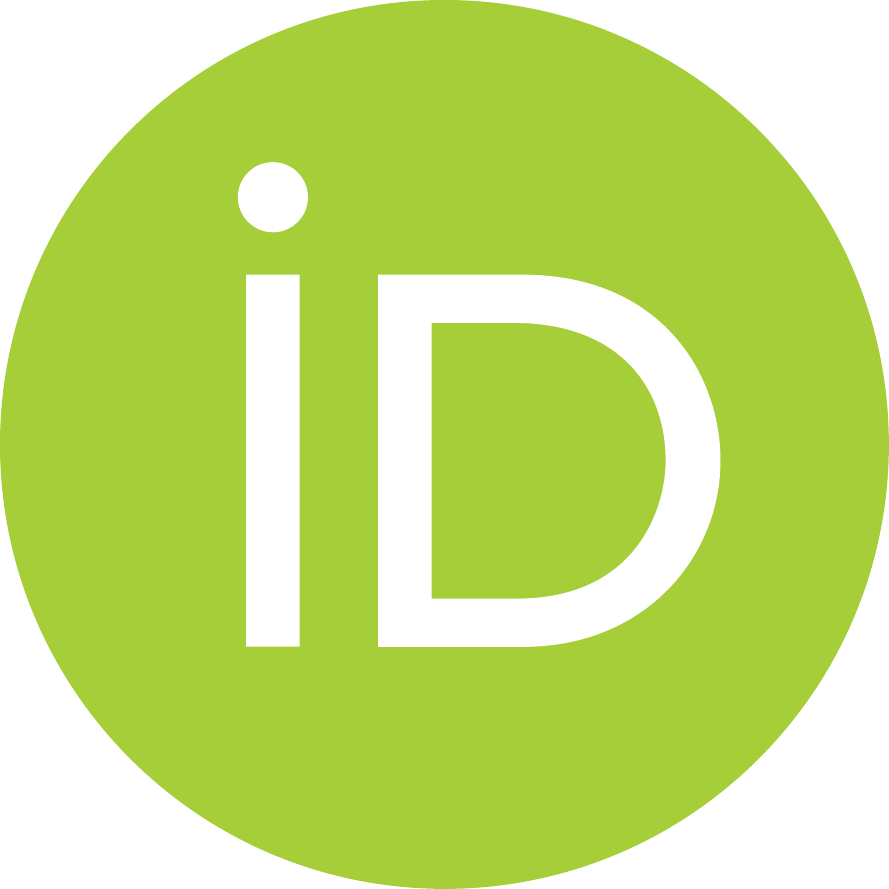}}}

\title{Performance Analysis of a Quantum Monte Carlo Application on Multiple Hardware Architectures Using the HPX Runtime}

\author{
Weile Wei\orcid{0000-0002-3065-4959} \\
  Louisiana State University\\
  \texttt{wwei9@lsu.edu} \\
   \And
Arghya Chatterjee\orcid{0000-0002-7259-2944} \\
  Oak Ridge National Laboratory\\
  \texttt{chatterjeea@ornl.gov} \\
  \And
Kevin Huck\orcid{0000-0001-7064-8417} \\
  University of Oregon\\
  \texttt{khuck@cs.uoregon.edu} \\
\And
Oscar Hernandez\orcid{0000-0002-5380-6951} \\
  Oak Ridge National Laboratory\\
  \texttt{oscar@ornl.gov} \\
\And
Hartmut Kaiser\orcid{0000-0002-8712-2806} \\
  Louisiana State University\\
  \texttt{hkaiser@cct.lsu.edu} \\
}

\begin{document}
\maketitle
\begin{abstract}
This paper describes how we successfully used the HPX programming model to port the DCA++ application on multiple architectures that include POWER9, x86, ARM v8, and NVIDIA GPUs. We describe the lessons we can learn from this experience as well as the benefits of enabling the HPX in the application to improve the CPU threading part of the code, which led to an overall 21\% improvement across architectures.
We also describe how we used HPX-APEX to raise the level of abstraction to understand performance issues and to identify tasking optimization opportunities in the code, and how these relate to CPU/GPU  utilization  counters,  device  memory  allocation over time, and CPU kernel level context switches on a given architecture. 
\end{abstract}


\footnotetext{This manuscript has been co-authored by UT-Battelle, LLC under Contract No. DE-AC05-00OR22725 with the U.S. Department of Energy.  The United States Government retains and the publisher, by accepting the article for publication, acknowledges that the United States Government retains a non-exclusive, paid-up, irrevocable, world-wide license to publish or reproduce the published form of this manuscript, or allow others to do so, for United States Government purposes. The Department of Energy will provide public access to these results of federally sponsored research in accordance with the DOE Public Access Plan (http://energy.gov/downloads/doe-public-access-plan).}

\section{Introduction}
\label{sec:Intro}



As users move their applications toward accelerated node architectures of different accelerator types and next-generation multi-core systems, they encounter significant challenges in their codes as there are few programming models available on all of these new architectures that can interoperate well with C++ and vendor specific APIs and libraries.
Our goal is to examine how successfully we can use the HPX programming model to port codes between architectures, and what lessons we can learn from this experience.
HPX also helps raise the level of abstraction in the application's programming model in order to understand common performance problems across architectures.
This helps to identify common optimization opportunities to hide latency, overheads, serializations and wait times while bringing performance improvements ``off-the-shelf'' to the application originally written using parallelism in C++.
In this paper, we explain which performance issues HPX can address and describe how we use it in the DCA++ application, its evaluation on different platforms, and how we can tune it to target to multiple platforms.
With rapidly changing configurations of highly heterogeneous HPC
systems, portability of code and performance of scientific applications is paramount for their software design and development efforts and long sustainability of applications. 

DCA++ (Dynamical Cluster Approximation) is a high-performance research software framework, providing a modern C++ implementation to solve quantum many-body problems \cite{PhysRevB.58.R7475, PhysRevB.61.12739, RevModPhys.77.1027}. The DCA++ code currently uses three different programming models (MPI, CUDA, and C++ Standard threads), together with numerical libraries (BLAS, LAPACK and MAGMA), to expose the parallelization in computations.

HPX is a C++ Standard Library for Concurrency and Parallelism \cite{hpx1, kaiser2009parallex,Kaiser:2015:HPL:2832241.2832244, Kaiser2020}. It implements all of the corresponding facilities as defined by the C++ Standard. Additionally, in HPX we implement functionalities proposed as part of the ongoing C++ standardization process.

In this paper, we outline HPX as a potential solution to efficiently porting DCA++ across different architectures. 

\subsection{Contribution}
The primary contributions of this work are outlined below:
\begin{enumerate}[label=(\alph*)]
    \item Ported DCA++ to various HPC architectures (POWER9, x86\_64, ARM64) (see \ref{sec:porting})
    \item Implemented the HPX threading model for on-node parallelization in DCA++
    \item Profiled DCA++ using performance measurement library APEX, integrated with HPX
    \item Collaborated with APEX performance observation tool team members, providing feedback and driving research
    \item Worked with DCA++ domain science application developers driving their new complex science problems with enhanced optimizations.
\end{enumerate}

\section{Background}
\label{background}



Quantum Monte Carlo (QMC) solver applications are common tools and mission critical
across the US Department of Energy's (DOE) application landscape. For the purpose 
of this manuscript the authors choose to use one of the leading QMC applications,
developed primarily at Oak Ridge National Laboratory in collaboration with ETH
Zúrich, the Dynamical Cluster Approximation (DCA++) algorithm. In recent
years DCA++ has been ported and successfully optimized across various platforms
(on both host side and accelerator based devices). A production scale scientific
problem runs on the DOE's fastest supercomputer, Summit, at Oak Ridge Leadership 
Facility (OLCF) on all 4600 nodes equipped with $\sim$28000 NVIDIA Volta V100 GPUs
attaining a peak performance of 73.5 PFLOPS with a mixed precision
implementation~\cite{DCA_pact19}. 

Although DCA++ has been higly optimized on existing hardware, this is the first effort to 
focus on the runtime execution level of the application and observe 
how it performs on each of the already supported systems and newer DOE supported
architectures. In this work, the authors enable HPX 
runtime support to further optimize thread context switching and lower
synchronization cost over the usage of C++ standard threads. We further verify such
claims using the APEX performance measurement tool.


\subsection{DCA++}

Dynamical Cluster Approximation (DCA++) is a numerical simulation tool that is 
used to predict behaviors of quantum materials, such as superconductivity, 
magnetism, etc. It is an iterative convergence algorithm with two primary kernels: 
\begin{inparaenum}[(a)]
\item Coarse-graining of the single-particle Green's function to reduce 
the complexity of the infinite size lattice problem to that of an effective 
finite size cluster problem, and,
\item Quantum Monte Carlo (QMC) based solver for the cluster problem.
\end{inparaenum}

\begin{figure}[ht]
	\centering
	\includegraphics[width=0.7\textwidth]{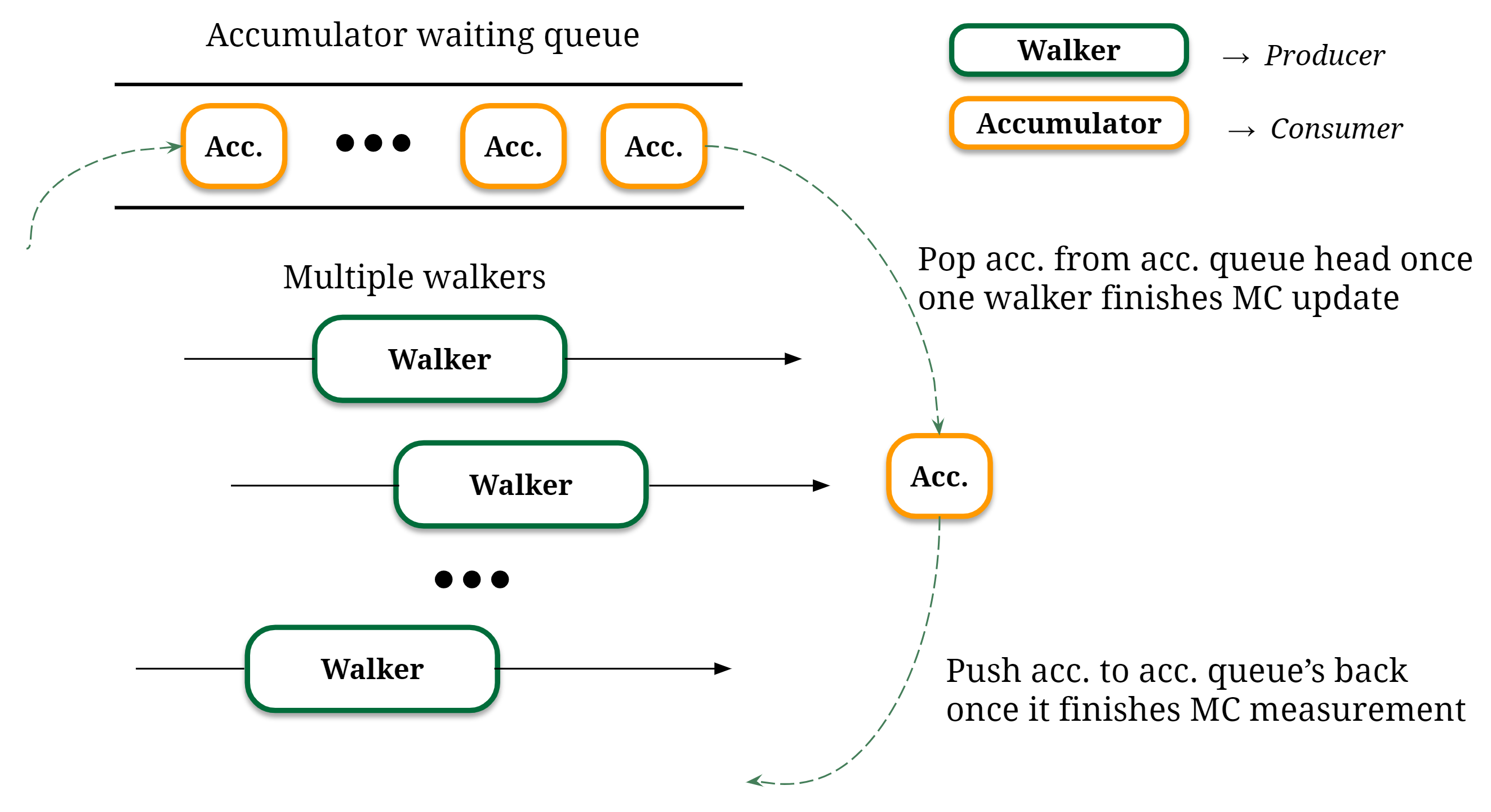}
	\caption{Shows the computation structure of a threaded QMC kernel using the custom-made thread pool in DCA++ running on a single MPI process (rank). We run multiple \emph{walker} threads concurrently, and after each \emph{walker} finishes an MC update, an idle \emph{accumulator} thread is pulled from the head of \emph{accumulator waiting queue} to compute MC measurement from the \emph{walker}. After the \emph{accumulator} finishes its measurement, it's pushed to the back of the queue.}
	\label{fig:threaded_qmc}
\end{figure}

Most of the application's performance, workload (computation), memory usage and 
bottlenecks come from the QMC solver kernel \cite{DCA_pact19}. Fig. 
\ref{fig:threaded_qmc} shows the on-node (per MPI process) computation structure of a
threaded QMC simulation using the custom-made thread pool in DCA++. We initialize 
several instances of independent Markov chains and distribute across nodes (MPI 
ranks), each node is responsible for that Markov chain assigned \footnote{On systems 
with the ability to run multiple MPI ranks per node with one or more GPUs per rank, 
each process is then only responsible for a portion of the chain assigned to that 
node}, computed by a \emph{walker object} (producer) and an \emph{accumulator 
object} (accumulator) that measures single- and two-particle Green's functions. 

Each object runs on an independent thread and no communication 
happens between these threads. We run multiple \emph{walker} threads concurrently, 
and after each \emph{walker} finishes a Monte Carlo (MC) update (sampling from the Markov
chain), the \emph{accumulator} is pulled from the head of \emph{accumulator waiting queue} to
compute MC measurement from the \emph{walker}. When each \emph{accumulator} finishes
its measurement, it's pushed into the back of the queue. The queries to the queue are
managed by the synchronization primitives (\lstinline{mutex} and
\lstinline{conditional_variable}).

In this paper the analysis, 
optimization, and further performance gains will be discussed in reference only to
the QMC solver portion of the DCA++ application.


\subsection{HPX}

HPX is a C++ standard library for distributed and parallel programming built on top of an asynchronous many-task runtime system (AMT).  It has been described in detail in other publications~\cite{heller:2012,Heller:2013:UHL:2530268.2530269,hpx_pgas_2014,Kaiser:2015:HPL:2832241.2832244,hartmut_kaiser_2020_598202,Heller2016}. Such AMT runtimes provide a means for helping programming models to fully exploit available parallelism on complex emerging HPC architectures. The HPX runtime includes the following essential components:

\begin{itemize}
\item An ISO C++ standard conforming API that enables wait-free asynchronous parallel programming, including {\it Futures}, {\it Channels}, and other primitives for asynchronous execution. The exposed API ensures syntactic and semantic equivalence of local and remote operations, which greatly simplifies writing complex applications~\cite{heller2019harnessing,daiss2019piz}.
\item A work-stealing lightweight task scheduler~\cite{kaiser2009parallex,hpx4} that enables finer-grained parallelization and synchronization, exposes greatly reduced overheads related to threading, and ensures automatic load balancing across all local compute resources (see~\ref{sec:hpx-threads}).
\item APEX~\cite{huck2015jsfi}, an {\it in-situ} profiling and adaptive tuning framework (see~\ref{sec:apex}).
\item In its distributed version (not utilized in the presented work), HPX also features an Active Global Address Space (AGAS)~\cite{hpx_pgas_2014,amini2019agas} that supports load balancing via object migration and enables runtime-adaptive data placement and distributed garbage collection and an active-message networking layer that enables running functions close to the objects they operate on~\cite{kaiser2009parallex,biddiscombe2017zero}. 
\end{itemize}
\smallskip

In the context of the presented work we use HPX because of its full conformance to the recent C++ standards\cite{standard2017programming, standard2020programming}, its reduced thread and synchronization overhead properties, and its sophisticated performance measurement and {\it in-situ} profiling capabilities provided by APEX.

\subsection{HPX-APEX Integration}
\label{sec:apex}

APEX~\cite{huck2015jsfi} (Autonomic Performance Environment for eXascale) is a performance measurement library for distributed, asynchronous multitasking runtime systems such as HPX. It provides support for both lightweight measurement and high concurrency. To support performance measurement in systems that employ user-level threading, APEX uses a dependency chain in addition to the call stack to produce traces and task dependency graphs. APEX supports both synchronous (so-called \textit{first person}) and asynchronous (\textit{third person}) measurements.
The synchronous module of APEX uses an event API and event listeners. Whenever an HPX task is created, started, yielded or stopped, APEX will respectively create, start/resume, yield, or stop timers for measurements. Dependencies between tasks are also tracked.
The asynchronous measurement involves periodic or on-demand interrogation of operating system, hardware or runtime states (e.g. CPU utilization, resident set size, memory ``high water mark'').  HPX counters (e.g. idle rate, queue lengths) are also captured on-demand on a periodic basis.


APEX has native support for performance profiling, in which all tasks scheduled by the runtime are measured and a report is output to disk and/or the screen at the end of execution. The profile data contains the number of times each task was executed and the total time spent executing that type of task. In order to perform detailed performance analysis involving synchronization and/or task dependency analysis, full event traces including event identification and start/stop times have to be captured. To that end, APEX is integrated with the Open Trace Format 2~\cite{eschweiler2011open} (OTF2) library -- an open, robust format for large scale parallel application event trace data. OTF2 is a robust reader/writer library and binary format specification that is typically used for high-performance computing (HPC) trace data. In order to capture full task dependency chains in HPX applications, all tasks are uniquely identified by their GUID (globally unique identifier) and the GUID of their parent task. These GUIDs are captured as part of the OTF2 trace output.  OTF2 data can be visualized by the Vampir~\cite{knupfer2008vampir} trace analysis tool.

Before the DCA+HPX integration, the \textit{first person} measurement in APEX was only integrated with a handful of technologies, incuding the HPX runtime and OpenMP 5.0 runtimes that support the OMPT performance tools interface~\cite{openmp50}.  The \textit{third person} measurement in APEX was mostly limited to extracting data from HPX and the Linux \texttt{/proc} virtual filesystem.   Because most of the DCA++ computation is offloaded to GPUs using the CUDA library, APEX was integrated with the CUDA Profiling Tools Interface (CUPTI)~\cite{cupti} and the NVIDIA Management Library (NVML)~\cite{nvml}.  Synchronous CUDA API callback timers and some counters (e.g. Bytes transferred, bandwidth, vector lanes) from the CUDA runtime and/or device API are captured synchronously, whereas the NVML counters (e.g. utilization, bandwidth, power) are periodically captured asynchronously.  Using APEX GUIDs mapped from CUDA Correlation IDs, the GPU activity such as memory transfers and kernel executions are captured and linked to the host-side tasks that launched them.  To provide concurrent use of the GPU hardware, memory transfers between the host and GPU and kernels are executed within logical subdivisions of the device, identified by the device, context, and stream IDs.  These IDs are associated with the OTF2 virtual ``threads'' of execution within the trace data, as shown in Fig.~\ref{fig:apex_timeline}.

\section{Implementation}
\label{section:implementations}

\begin{figure}[ht]
	\centering
	\includegraphics[width=0.7\textwidth]{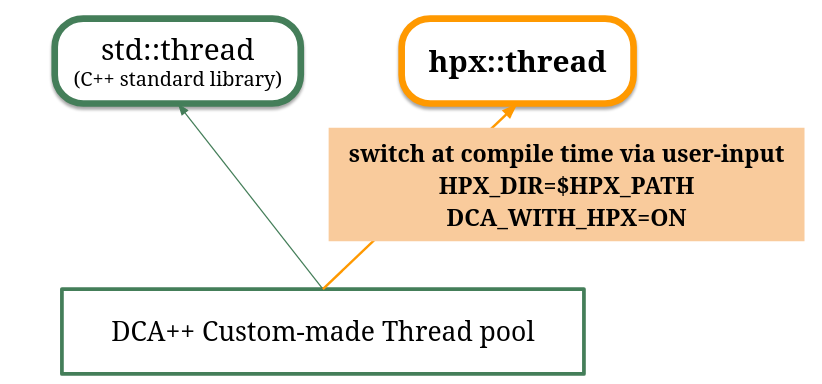}
	\caption{Custom-made thread pool in DCA++ now supports both \lstinline{std::thread} (default) and \lstinline{hpx::thread} (new feature). Threading options can be toggled at compilation.}
	\label{fig:threading_abtraction_layer}
\end{figure}

In this section, we outline our implementation of the high-level 
threading abstraction layer in DCA++, which supports standard C++
threading and HPX threading implementations\footnote{\url{https://github.com/STEllAR-GROUP/DCA/releases/tag/hpx_thread}}. 
The design of HPX integration in DCA++ is presented in Fig.~
\ref{fig:threading_abtraction_layer}. Our implementation is non-intrusive to DCA++ 
code as it does not break the API of the custom-made thread pool and we have not 
modified original DCA++ workflow. It also allows the application developer to switch 
between \lstinline{hpx::thread} and \lstinline{std::thread} via compilation 
configuration. If user prefers HPX threading option, one needs to turn on 
\lstinline{DCA_WITH_HPX} flag and provide the path of HPX library to the 
application's CMake configuration.


To parallelize computation tasks, DCA version 1.1.0\footnote{\url{https://github.com/CompFUSE/DCA/releases/tag/paper.2019.old_code}} implemented  a multi-threading strategy using POSIX threads which
could cause large overheads when thousands of threads continuously spawned and joined.
DCA version 2.0\footnote{\url{https://github.com/CompFUSE/DCA/releases/tag/paper.2019.new_code}} lowered the overhead with the custom-made thread pool strategy \cite{DCA_pact19} by maintaining
constant number of C++ \lstinline{std::thread} objects during the execution. However, the implementation
of the custom-made thread pool strategy was designed to spread worker threads to
simultaneous multithreading (SMT) or virtual cores. Depending on the architecture
of the processor, SMT might be a bottleneck if any of the SMT threads are competing
for the shared physical core \cite{saini2011impact}.

We manage to preserve the same API of the \lstinline{ThreadPool} implementation in both versions primarily due to the fact that HPX is fully C++ standard conforming.
All synchronization primitives of the standard C++ library are still valid in the context of HPX. For the C++ \lstinline{std::thread} version of the thread pool shown in Listing \ref{std_class}, we wrapped all C++ standard synchronization primitives (i.e. \lstinline{condition_variable}, \lstinline{lock_guard}, \lstinline{future}) into a \lstinline{thread_traits} class. 
For the HPX-enabled DCA++ shown in Listing \ref{hpx_class}, we construct a similar \lstinline{thread_traits} class in a separate header file and replace all the C++ standard synchronization primitives with equivalent HPX synchronization primitives.

\begin{lstlisting}[language=C++, caption={\lstinline{std::thread} version of the thread pool.}, label={std_class}]
namespace dca { namespace parallel {
struct thread_traits {
    template <typename T>
    using future_type = std::future<T>;
    using mutex_type = std::mutex;
    using condition_variable_type = std::condition_variable;
    using scoped_lock = std::lock_guard<mutex_type>;
    using unique_lock = std::unique_lock<mutex_type>;
};

class ThreadPool {...};
}}
\end{lstlisting}

\begin{lstlisting}[language=C++, caption={\lstinline{hpx::thread} version of the thread pool. Note that for the synchronization primitives implemented in \lstinline{class thread_traits}, this version differs from the \lstinline{std::thread} version only by the used C++ \lstinline{namespace hpx}.}, label={hpx_class}]
namespace dca { namespace parallel {
struct thread_traits {
    template <typename T>
    using future_type = hpx::future<T>;
    using mutex_type = hpx::mutex;
    using condition_variable_type = hpx::condition_variable;
    using scoped_lock  = std::lock_guard<mutex_type>;
    using unique_lock = std::unique_lock<mutex_type>;
};

class ThreadPool {...};
}}
\end{lstlisting}

For task-scheduling in the custom-made thread pool implemented in \lstinline{class ThreadPool}, the C++ \lstinline{std::thread} version of the thread pool \cite{DCA_pact19} maintains an array of \lstinline{std::thread} objects and array of
queues of work items represented by \lstinline{std::packaged_task} objects in a simple round-robin fashion; 
HPX threading version dispatches tasks asynchronously through \lstinline{hpx::async} and
manages tasks with its runtime scheduler that has various robust task scheduling methods \cite{zhang2019introduction}.

For thread affinity, the C++ \lstinline{std::thread} version of the thread pool 
manually sets thread affinity and uses the (SMT) feature to achieve speedup 
\cite{DCA_pact19}; the \lstinline{hpx::thread} version on the other hand handles
these scheduling efforts automatically through its runtime system. 
HPX by default recognizes existing SMT and sets only one hyper-thread per physical 
processing unit. The runtime schedules user-level lightweight threads on top of operating system threads, which avoids expensive context switches at kernel-level \cite{hpx4}.

\label{sec:hpx-threads}
HPX-threads are implemented as user-level threads. These are cooperatively (non-preemptively) scheduled in user mode by the HPX-thread manager on top of one OS thread per hardware thread (processing unit). By default, the OS threads have their affinities defined such that they run on one processing unit only. The HPX-threads can be scheduled without a kernel transition, which provides a performance boost. Additionally, the full use of the OS’s time quantum per OS-thread is achieved even if an HPX-thread blocks for any reason.  In that case, other HPX-threads are scheduled to run immediately. The scheduler is cooperative in the sense that it will not preempt a running HPX-thread until it finishes execution or cooperatively yields its execution. This is particularly important, since it avoids context switches and cache thrashing due to randomization introduced by preemption. The default thread scheduler is implemented as a ‘First Come First Served’ scheduler, where each OS-thread works from its own queue of HPX-threads. Other scheduling policies, e.g. supporting thread priorities, are available as well. If one of the cores runs out of work, it starts ‘stealing' queued tasks from neighboring cores, thus enabling load-balancing across all cores~\cite{kaiser2009parallex,hpx4}.

\section{Results}
\label{section:results}
\newcommand{\kernel}[1]{\textit{#1}}

\subsection{Systems overview}

For our evaluation, we have used Oak Ridge Leadership Computing Facility's
(OLCF) Summit supercomputer and the Wombat system; and, National
Energy Research Scientific Computing Center's (NERSC) Cori Supercomputer (for this
work we used the new CoriGPU partition). 
Each system was selected due to its host architecture diversity
(shown in Table. \ref{tab:systems_comparison}) for comparing the performance of DCA++ using the HPX runtime and visualizing the results collected using APEX and visualized by Vampir. 

{\bf Summit.} \cite{summit_guide}
is a 4600 node, 200 PFLOPS IBM AC922 system \footnote{Summit ranked the second place in the TOP500 list in June 2020 \cite{TOP500}}. Each node consists of 2 \emph{IBM
POWER9} CPUs with 512 GB DDR4 RAM and 6 NVIDIA V100 GPUs with total of 96 GB
high bandwidth memory (divided into 2 sockets), all connected together with
NVIDIA's high-speed NVLink. 

\begin{table*}[ht]
\begin{center}
\caption{Systems Comparison}
\label{tab:systems_comparison}
{\renewcommand{\arraystretch}{1.3}
\begin{tabular}{|p{0.15\textwidth}|p{0.25\textwidth}|p{0.25\textwidth}|p{0.25\textwidth}|}
\hline
\bf{Configuration} & \bf{Summit} & \bf{Wombat} & \bf{CoriGPU} \\ \hline
GPU & NVIDIA Volta (6 per node) & NVIDIA Volta (2 per node) & NVIDIA Volta (8 per node) \\ \hline

\multirow{2}{*}{CPU}
& IBM POWER9™ (2 Sockets / 21 Cores per socket) 
& Cavium ThunderX2 (2 Sockets / 28 Cores per socket) 
& Intel Xeon Gold 6148 (2 sockets / 20 cores per socket)  \\ \hline

CPU-GPU interconnect
& NVIDIA NVLINK2 (50 GB/s)
& PCIe Gen3 (16 GB/s) 
& PCIe Gen3 (16 GB/s) \\ \hline

\end{tabular} }
\end{center}
\vspace{-.2in}
\end{table*}

{\bf Wombat.} \cite{wombat_guide}
is a 64-bit ARM cluster with 16 compute nodes, four of which have two NVIDIA V100 GPUs attached.
Each compute node has two 28-core \emph{Cavium ThunderX2 processors} (Cavium is now Marvell),
256 GB RAM (16 DDR4 DIMM’s) and a 480 GB SSD for node-local storage. Nodes are connected with EDR 
InfiniBand ($\sim$100 Gbit/s).

{\bf CoriGPU.} \cite{corigpu_guide} is a 
development rack of 18 nodes recently added to the Cori system at NERSC. 
Each node contains two 20-core \emph{Intel Xeon Gold 6148} CPUs 
with 384GB DDR4 memory and 8 NVIDIA V100 GPUs with 128 GB HBM2 memory 
(divided into 2 sockets). All GPUs are connected to the CPUs and Infiniband network 
interface cards via PCIe 3.0.

\subsection{Correctness verification across systems} \label{sec:porting}

To verify the correctness of our work across various HPC architectures, 
we follow the standard DCA++ protocol\footnote{\url{https://github.com/CompFUSE/DCA/wiki/Tutorial:-Tc}} 
to study superconductivity in the 2D single-band Hubbard model in DCA++. The focus value is the
superconducting transition temperature $T_c$, a property of the materials.
We choose 100k Monte Carlo measurements as it is representative case to our science problems.
The goal is to obtain the same $T_c$ with acceptable statistical noise across all HPC architectures for a specific scientific case as defined under the protocol.  

\begin{figure}[H]
\centering
\subfloat[Here we validate our science case with C++ \lstinline{std::thread} 
implementation across three HPC platforms.]
    {\includegraphics[width=0.7\textwidth, trim=0cm 0cm 0cm 1.2cm,clip]{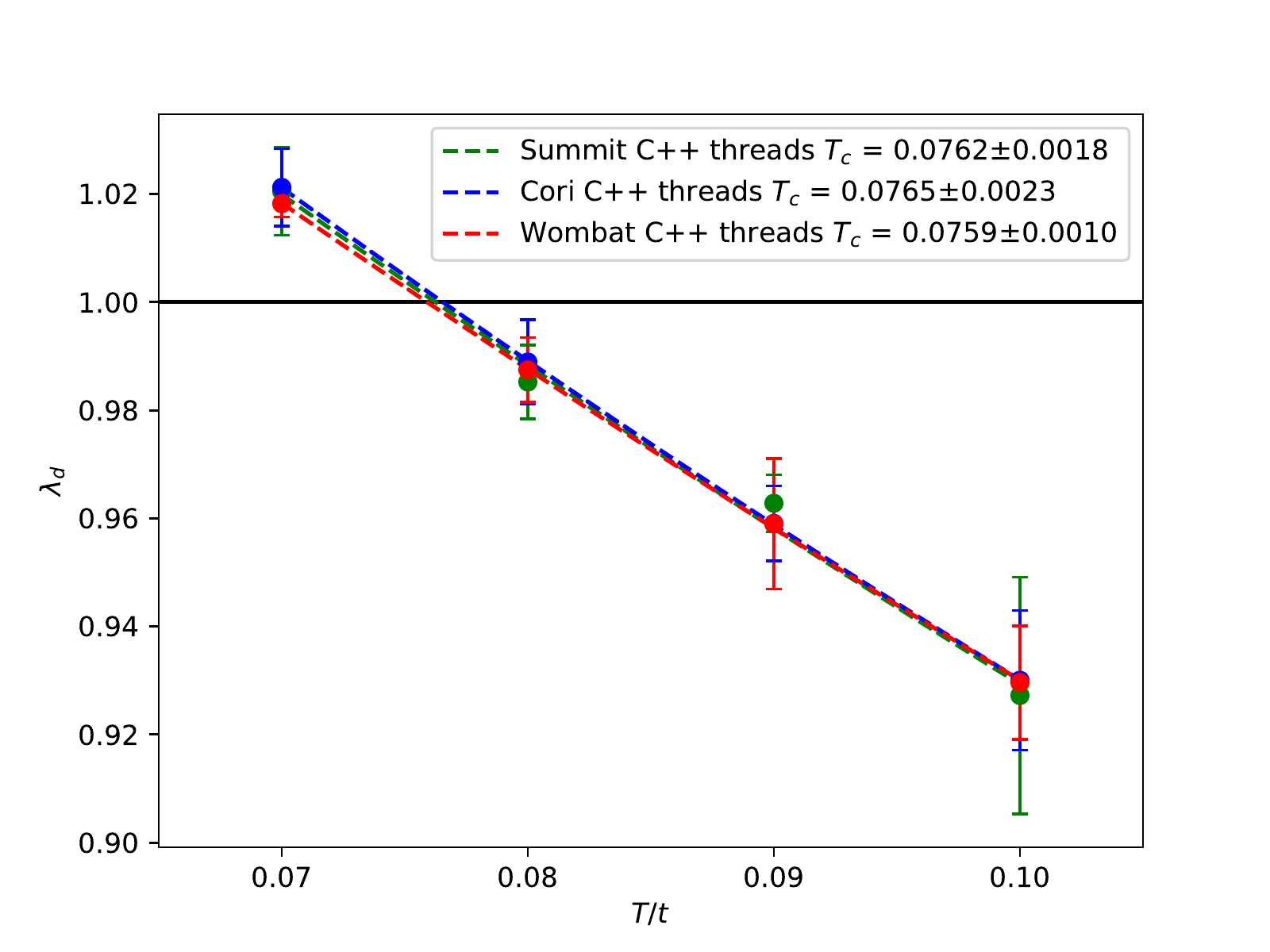}\label{correctness_test_std}}

\subfloat[Validation using the same case with \lstinline{hpx::thread} implementation across the same three systems. Additionally, we show the C++ \lstinline{std::thread} results on Summit as a reference.]
    {\includegraphics[width=0.7\textwidth, trim=0cm 0cm 0cm 1.2cm,clip]{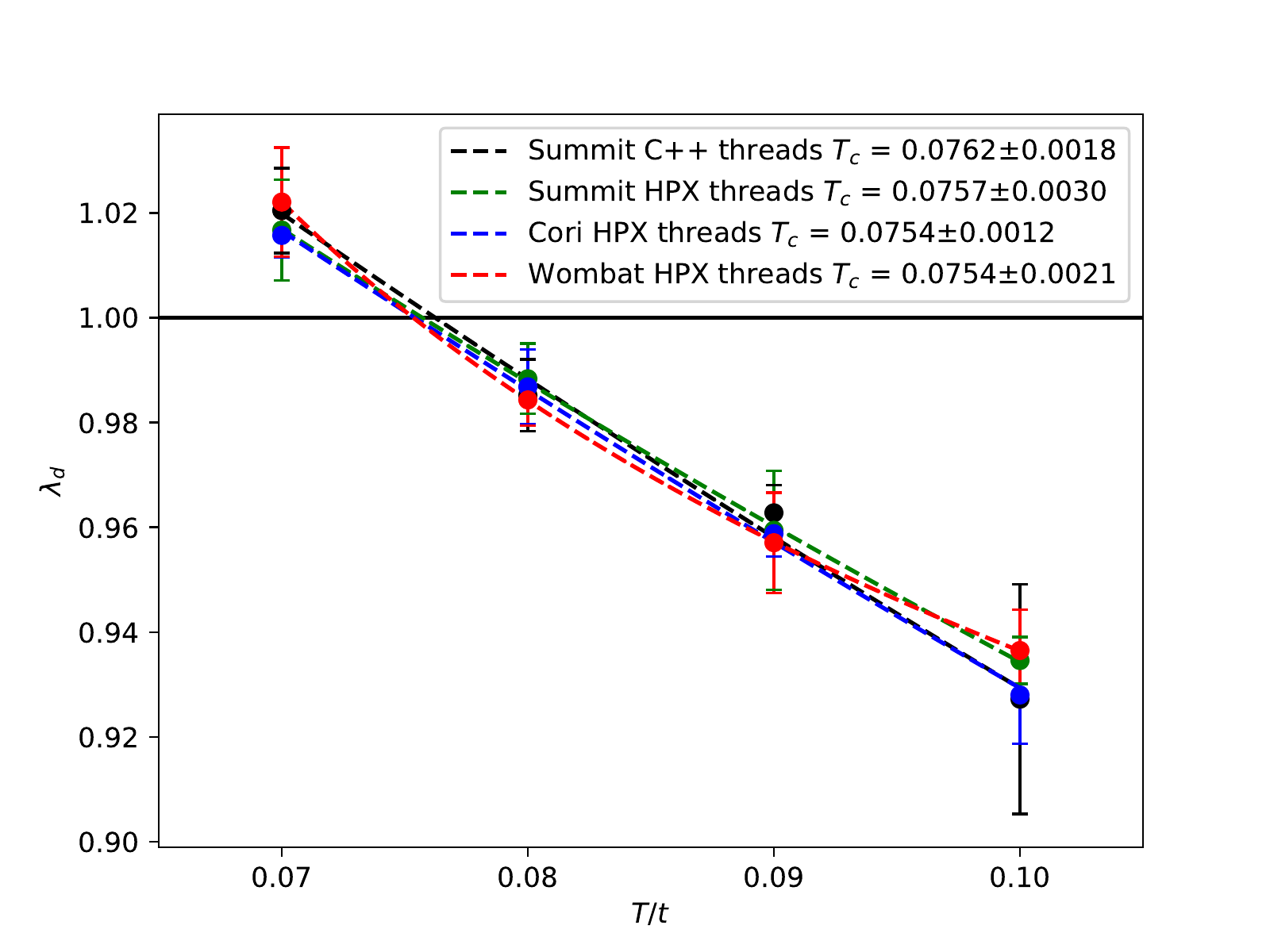}\label{correctness_test_hpx}}

\caption{DCA++ correctness verification across multiple architectures as outlined in Table~\ref{tab:systems_comparison}. For our scientific problem we obtain same
    superconducting transition temperature $T_c$ results 
	(where leading eigenvalue $\lambda_d$(T=$T_c$) = 1) within acceptable statistical range. For each platform, we compute DCA++ with 100k Monte 
	Carlo measurements (representative case to our science problems) for 5 independent calculations. The random number generator used in all experiments is \lstinline{std::mt19937_64} from C++ library.}
\label{fig:hpx_std_correctness} 
\vspace{-.2in}
\end{figure}

Fig. \ref{correctness_test_std} shows DCA++ with C++ \lstinline{std::thread} threading 
generates consistent results across various platforms. It
shows the temperature dependence of the leading eigenvalue $\lambda_d$ of the Bether-Salpeter equation. 
$T_c$ is the temperature where $\lambda_d$(T=$T_c$) = 1. All $T_c$ are about 0.076 within acceptable statistical range. 
Similarly, Fig. \ref{correctness_test_hpx} shows DCA++ with \lstinline{hpx::thread} also
generates accurate results across multiple HPC architectures. 
We use the DCA++ application with C++ \lstinline{std::thread} threading results obtained from
runs on Summit as a referencing result, and compare with all other runs of DCA++ using \lstinline{hpx::thread} on various platforms. 
As one might note that we have obtained the same $T_c$ within an acceptable statistical deviation.

\subsection{Compare runtime: \textit{std::thread} v.s. \textit{hpx::thread}}

For this comparison analysis we compared a version of DCA++ with C++ \lstinline{std::thread} and one with a \lstinline{hpx::thread} 
implementation on a single Summit node with 6 MPI ranks, each rank mapped to 7 physical cores and 1 Volta V100 GPU. More performance analysis (i.e. performance analysis on other machines) will be uploaded to the public repository \footnote{\url{https://github.com/STEllAR-GROUP/dca}} once available.

\begin{figure}[H]
	\centering
	\includegraphics[width=0.7\textwidth]{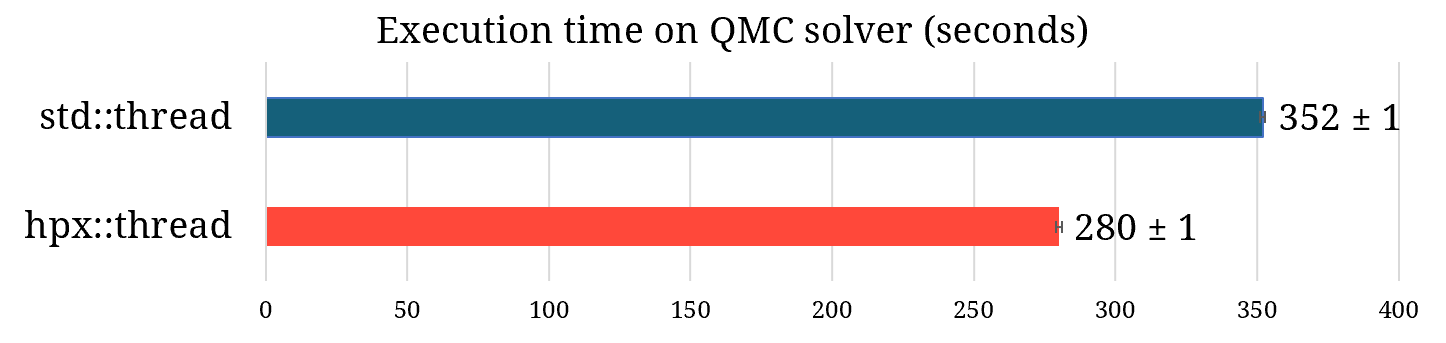}
	\caption{Time-to-solution for 100k Monte Carlo measurements 
	with error bars obtained from 5 independent executions on Summit.
	 Using the \lstinline{hpx::thread} implementation we observe up to 
	 21\% speedup over the  C++ \lstinline{std::thread} version.
	 Observed performance gain is due to faster context switch and scheduler and less synchronization overhead in HPX runtime system. Lower is better.}
	\label{fig:hpx_vs_std}
\vspace{-.2in}
\end{figure}

\begin{figure}[H]
	\centering
	\includegraphics[width=0.7\textwidth, trim=0cm 0cm 0cm 0.5cm,clip]{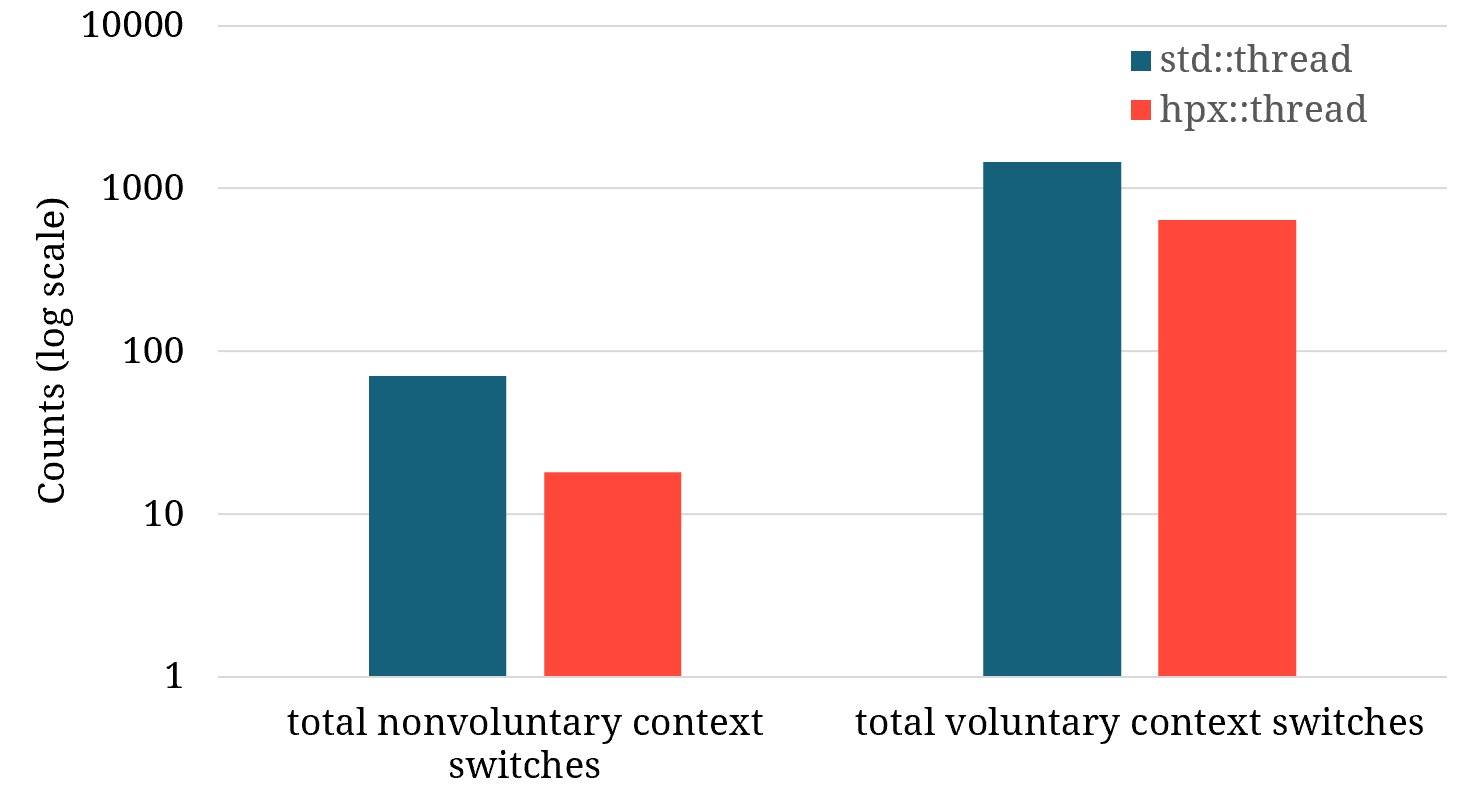}
	\caption{Comparison of non-voluntary and voluntary context switches 
	using the APEX performance measurement counters when executing DCA++ with C++ \lstinline{std::thread} and \lstinline{hpx::thread} versions on Summit. 
	We observe that the \lstinline{hpx::thread} implementation has much less context switches than \lstinline{std::thread} in DCA++ and aides to the 
	performance gains in using HPX over \lstinline{std::thread}. Lower is better.
 }
	\label{fig:apex_count_hpx_vs_std}
\vspace{-.2in}
\end{figure}

Fig. \ref{fig:hpx_vs_std} shows DCA++ with \lstinline{hpx::thread} achieves 21\% speedup over the one with C++ \lstinline{std::thread} version. The same improvement is also observed in the distributed runs as well. The speedup is mainly due to faster thread context switching and reduced scheduler and synchronization overheads in the HPX
runtime system (see Section~\ref{sec:hpx-threads}). Fig. \ref{fig:apex_count_hpx_vs_std}
verifies the speedup and shows by the end of the execution, \lstinline{hpx::thread}
version has much less ($\sim 2\times$ lower) voluntary context switches (639 times)
relative to \lstinline{std::thread} version (1454 times) and $\sim 4\times$ lower
non-voluntary context switches (18 times) relative to \lstinline{std::thread} version (70 times). For the non-voluntary context switches observed in \lstinline{hpx::thread} version, we consider these are most likely caused by the synchronization introduced by CUDA itself as
CUDA synchronization is still happening on pthread level.

\begin{figure}[H]
	\centering
	\includegraphics[width=0.7\textwidth]{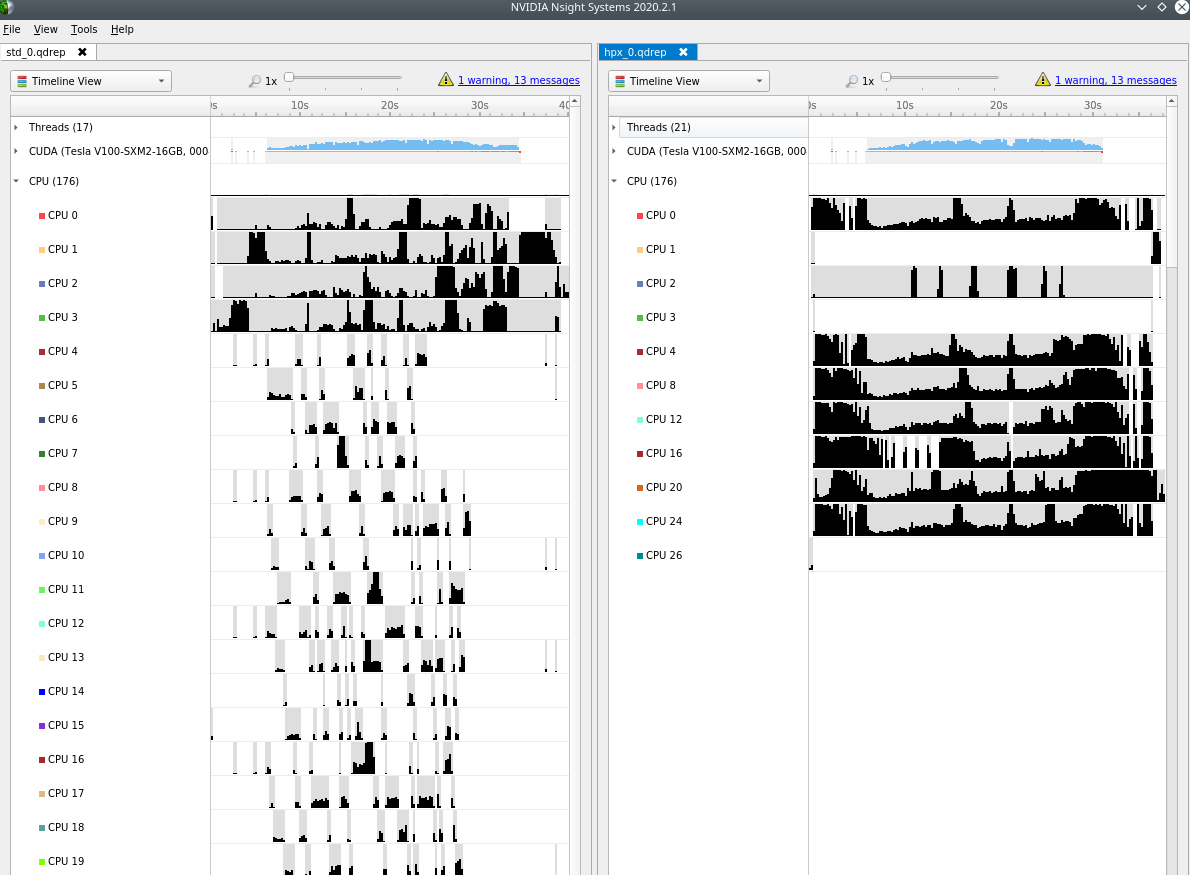}
	\caption{NVIDIA Nsight System profiler showing CPU utilization; for 
	C++ \lstinline{std::thread} (left, shows only 28 active hyper-threads) and \lstinline{hpx::thread} (right) versions for DCA++. 
	\lstinline{hpx::thread} version sets one hyper-thread per physical core to achieve better
	hardware utlization while \lstinline{std::thread} spreads work over 4 hyper-threads per physical core.}
	\label{fig:nsight_thread_affinity}
\end{figure}

Fig. \ref{fig:nsight_thread_affinity} was generated using the NVIDIA Nsight Systems on
Summit. The figure shows two different threading affinity strategies adapted in C++ 
\lstinline{std::thread} (left) and \lstinline{hpx::thread} version (right) in DCA++. 
Each row in the figure represents average hardware thread utilization. The height of the 
hardware thread utilization is represented by the height of the black histogram. 

For our test case we set the SMT to 4 for both executions. The C++ 
\lstinline{std::thread} version uses 4 hardware threads per physical core; while, 
HPX-enabled DCA++ by default utilizes only one hardware thread per physical core. Also, 
if we combine the adjacent 4 hyper-threads (SMT) for each physical core in C++ 
\lstinline{std::thread} version, the overall utilization is not as high as in the 
\lstinline{hpx::thread} version. Moreover, even if the DCA++ is modified to use the same
affinity settings (which requires explicit changes in the code base) as HPX, the 
performance is not improved (i.e. the affinity settings do not cause the speedup of 
HPX). The reason of the speedup is due to the fact that HPX thread management (and 
context switching in particular) exposes less overheads and lower synchronization 
overheads. With faster context switch from HPX threads, DCA++ is able to feeds more computing workload into GPU faster. This directly increases the GPU utilization resulting in the observed performance improvement. 

We further verified that thread caching malloc (i.e. tcmalloc) is not the cause of the speedup with
\lstinline{hpx::thread} version which uses tcmalloc. TCMalloc assigns each thread a thread-local cache and reduces lock contention for multi-threaded programs\cite{ghemawat2009tcmalloc}. 
We performed \lstinline{LD_PRELOAD tcmalloc} for DCA++
\lstinline{std::thread} version, and the execution time remains the same as the one without tcmalloc. This finding strengthens our conclusion that the 21\% speedup seen for the  \lstinline{hpx::thread} version is due to the fact that user-level context switching is more efficient and synchronization with HPX threads imposes less overhead (see Fig.~\ref{fig:hpx_vs_std}).

\subsection{HPX-APEX Profiling Analysis}

APEX was originally designed to be integrated with the HPX runtime, and enabling APEX support is straightforward. When configuring HPX, flags are passed to CMake in order to enable APEX support and provide the path to library dependencies such as OTF2, CUPTI and NVML. After configuration, build and installation the HPX runtime will have APEX performance measurement enabled. As mentioned in Section~\ref{background}, all HPX tasks are timed by APEX. In addition, tasks defined in the application can be annotated to provide unique labels using the \lstinline{hpx::annotated_function} facility in HPX. At runtime, different APEX features (e.g. tracing, output summary format, different counter sets) are enabled/disabled through the use of environment variables, a configuration file, or the APEX programming interface. 

For the experiments described below, APEX collected a full event trace to OTF2 and monitored several HPX, operating system, CPU and GPU utilization counters. 
Counters that were particularly useful for these experiments include kernel-level context switches (both voluntary and not), user and system level CPU utilization, GPU utilization and memory consumption, HPX idle rates and queue lengths.

\begin{figure}[!t]
\centering
\subfloat[Master timeline plot monitored events including CPU and GPU activities]{\includegraphics[width=0.7\textwidth]{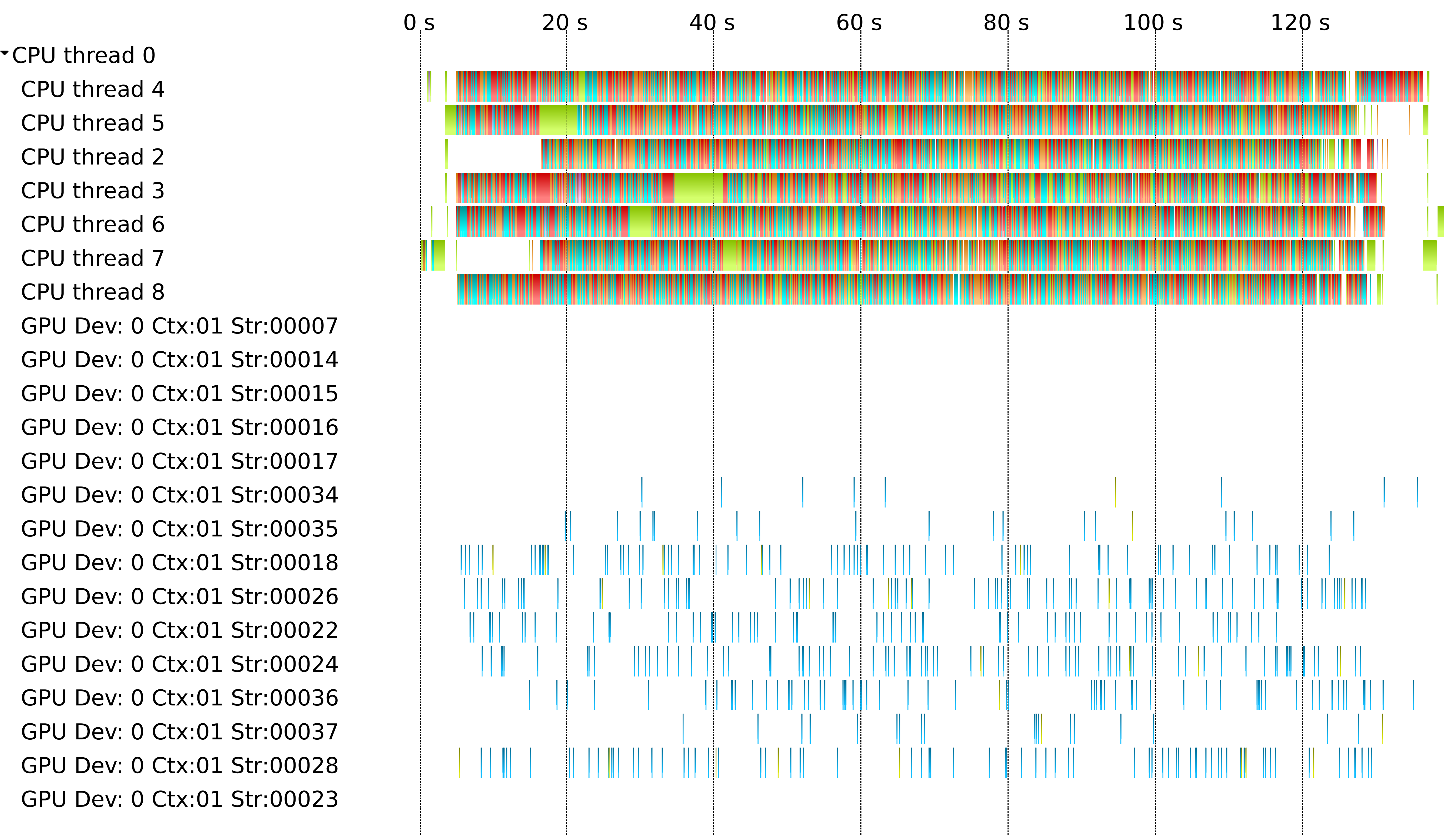}}\label{timeline}

\subfloat[Top 10 time consuming functions. Both annotated functions (user defined kernels) and CUDA API calls can be captured. Exclusive time means the amount of time spent in just this function and no subroutines are included.]
    {\includegraphics[width=0.7\textwidth, trim=0cm 2.3cm 0cm 0cm,clip]{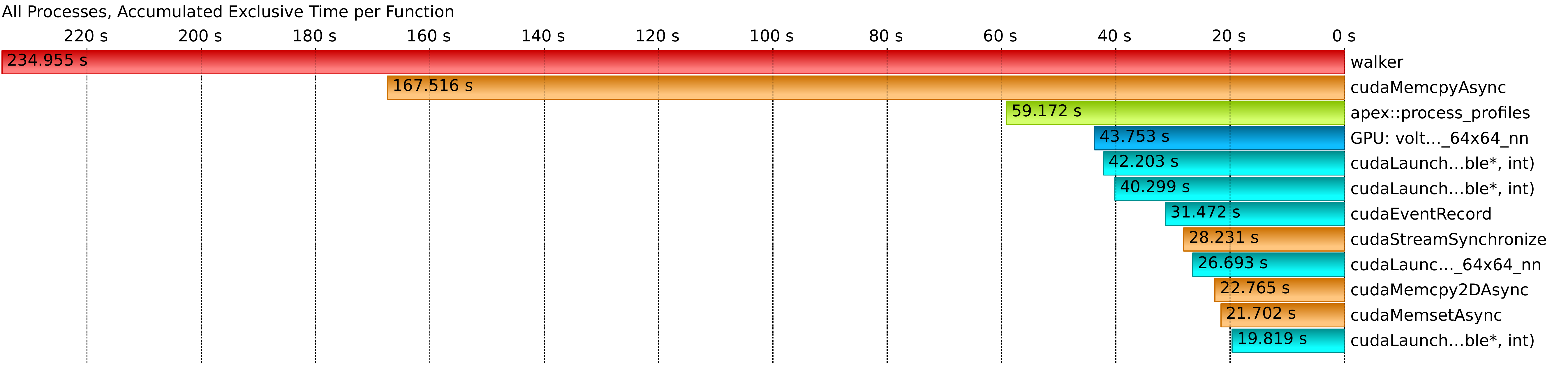} \label{apex_function_summary}}

\caption{HPX-APEX profiling results on Summit summarizing CPU and GPU activities.}
\label{fig:apex_timeline} 
\vspace{-1\baselineskip}
\end{figure} 

\begin{figure}
\centering
\subfloat[HPX thread idle rate (unit: 0.01\%), lower is better.]
{\includegraphics[width=0.7\textwidth, height=2.5cm, trim=6cm 0cm 0cm 0cm,clip]
{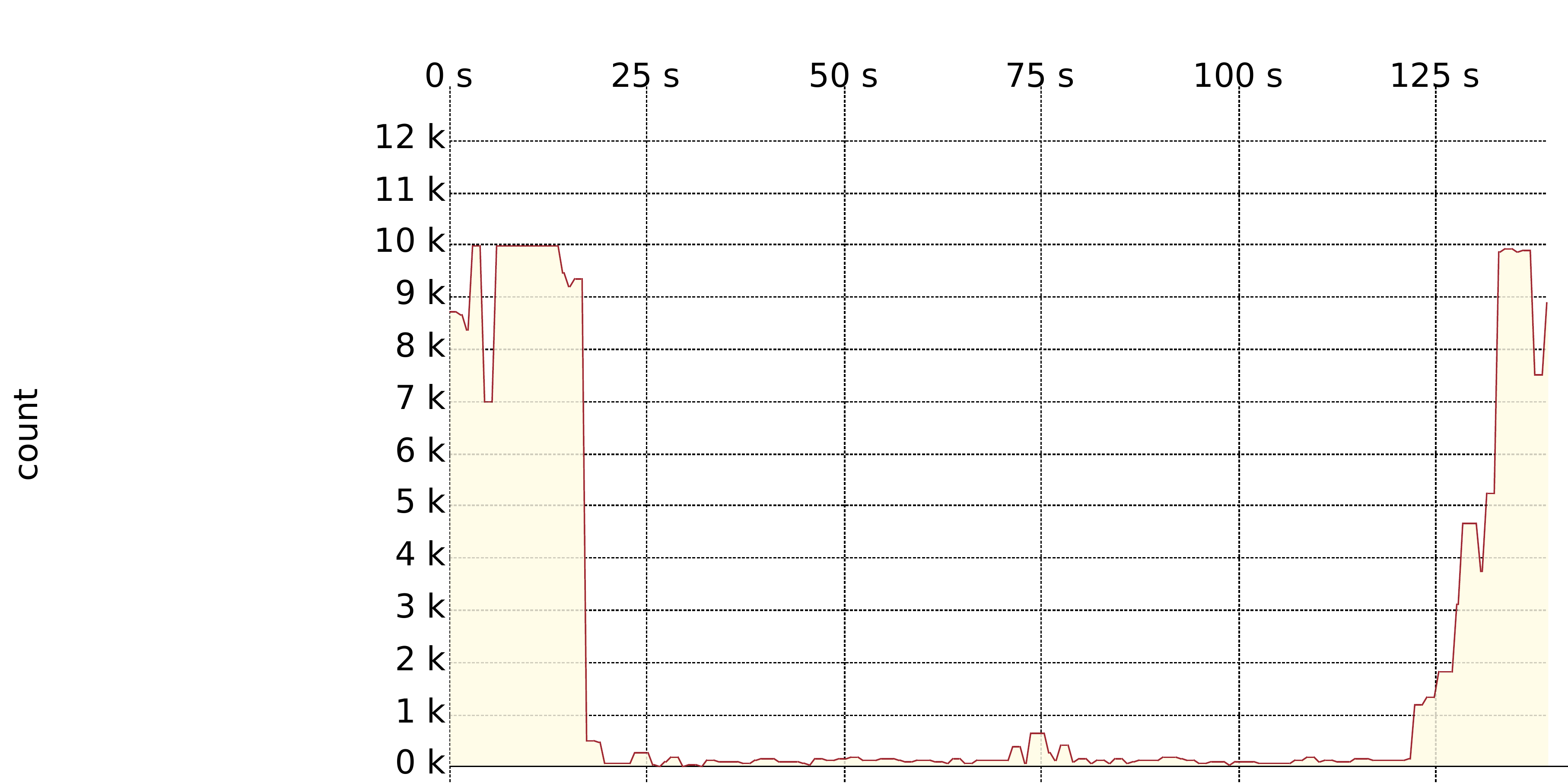} \label{hpx_idle_rate}} 

\subfloat[HPX queue length (unit: number of available tasks), higher is better.]
{\includegraphics[width=0.7\textwidth, height=2.5cm, trim=6cm 0cm 0cm 0cm,clip]
{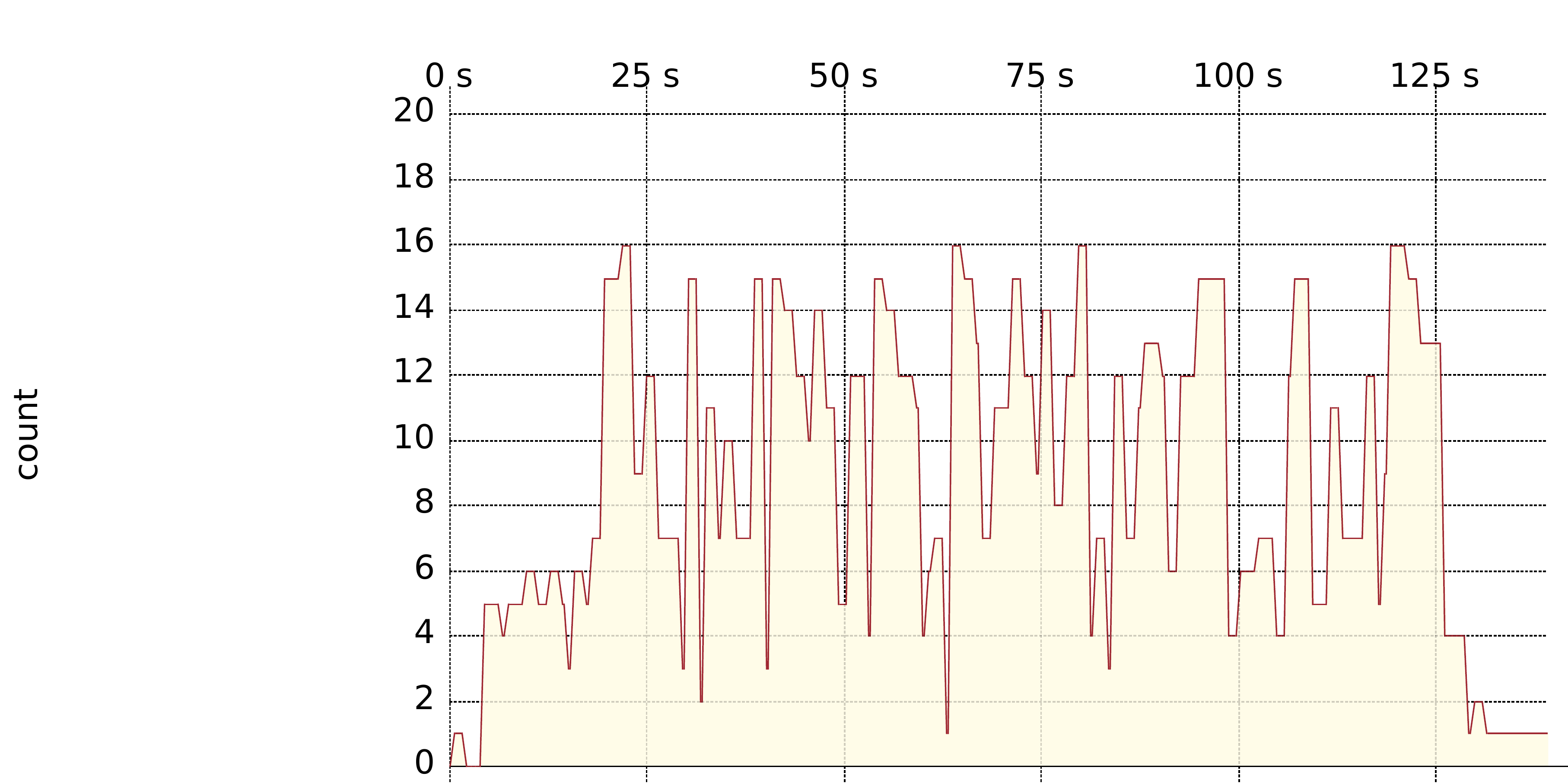} \label{hpx_queue_length}}
\caption{HPX-APEX profiling on Summit showing HPX thread idle rate and queue length.} 
\label{fig:apex_hpx}
\vspace{-1\baselineskip}
\end{figure}

We traced DCA++ with APEX on Summit as shown in Fig. \ref{fig:apex_timeline}. We are able to annotate any functions
with \lstinline{hpx::annotated_function} function wrapper in the code to distinguish their execution time
in final profiling data. Here we annotate \texttt{walker} and \texttt{accumulator} functions, as they are the most 
computation-intensive parts in DCA++ code. From Fig. \ref{apex_function_summary}, one can clearly observe that the walker function
takes majority of the time in a single DCA++ run. The profiling measurement library can also gather HPX thread idle rate (as seen in Fig.~\ref{hpx_idle_rate}) and queue length (as seen in Fig.~\ref{hpx_queue_length}). 
The idle rate counter indicates how utilized each of the HPX worker threads are during each sampled time period (lower is better). In the context of HPX, it is not a problem having the shown queue lengths as creating and managing HPX threads (tasks) is generally very cheap (less than 1 $\mu$s per thread). The queue depth indicates how much work, in the form of queued tasks, is available for each of the worker threads. The counters are collected on a per-worker basis, and the values shown here represent averages across all worker threads.

\begin{figure}[]
	\centering
	\includegraphics[width=0.7\textwidth, height=2.5cm, trim=6cm 0cm 0cm 0cm,clip]{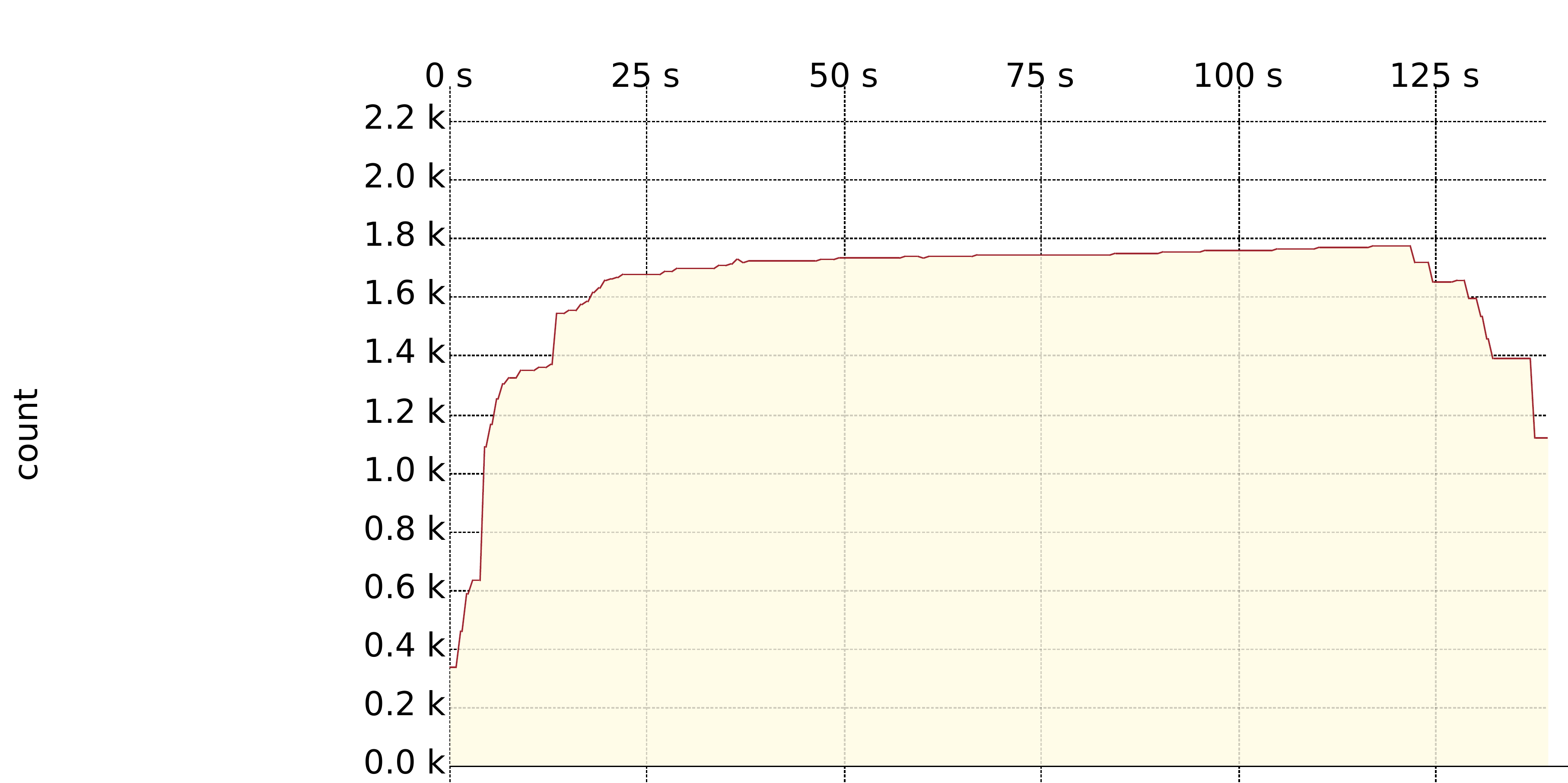}
	\caption{HPX-APEX profiling results on Summit summarizing device memory used (unit: megabyte) over the time.}
	\label{fig:apex_cudamalloc}
\vspace{-1\baselineskip}
\end{figure}

In~\cite{DCA_pact19} authors reported that while storing two-particle green function 
\textit{$G_{tp}$} on the device allows condensed matter physicists to explore larger 
and more complex (higher fidelity) science problems, 
but we are limited to the device memory size. The data of device memory usage from HPX-APEX
shown in Fig. \ref{fig:apex_cudamalloc} can help us track memory usage and provide computational scientists guidance on how to address
memory-bound challenge as defined in~\cite{DCA_pact19}. We are planning to distribute \textit{$G_{tp}$} across nodes and implement a token ring algorithm to transfer single-particle Green’s function \textit{$G$} between nodes. The implementation will take advantage of high-speed network between devices available on the machine (i.e. NVIDIA NVLink on Summit) in order to transfer device data efficiently.

\section{Conclusion}
\label{concl}

In this paper we used the Dynamical Cluster Approximation (DCA++) one of the leading Quantum Monte Carlo solvers as a research vehicle to test the feasibility of the HPX runtime system and use the abstraction layer in the programming model to understand the performance bottlenecks across multiple architectures (both host side and accelerator based devices). 

We observed significant performance benefit ($\sim$21\% speedup over standard threads) by just using the HPX threading model due to the faster context switches and lower synchronization overheads guaranteed by the HPX runtime. In this work we also validated our claims using the APEX performance measurement library and with the HPX-APEX integration one can observe in-depth analysis of the threading behavior (eg. CPU / GPU utilization counters, device memory allocation over time, kernel level context switches and more). 




\section{Acknowledgment}
\label{Ack}

Authors would like to thank Thomas Maier (ORNL), Giovanni Balduzzi (ETH Zurich) and Ed D'Azevedo (ORNL) for their
insights during the optimization phase of DCA++. The authors thank John Biddiscombe (ETHZ / CSCS) for initiating the port of
DCA++ to HPX, for providing the initial implementation, and insightful
discussions. OLCF system admins Matt Belhorn (ORNL) and
Ross Miller (ORNL) assisted the authors with setting up libraries / programming models on 
Summit and Wombat. We would further like to thank Marc Day and Kevin Gott (LBNL) for
assisting us with allocation on the CoriGPU at NERSC.

This work was supported by the Scientific Discovery through Advanced Computing (SciDAC) 
program funded by U.S. Department of Energy, Office of Science, Advanced Scientific Computing 
Research (ASCR) and Basic Energy Sciences (BES) Division of Materials Sciences and 
Engineering, as well as the RAPIDS SciDAC Institute for Computer Science and Data
under subcontract 4000159855 from ORNL.
This research used resources of the Oak Ridge Leadership Computing Facility, 
which is a DOE Office of Science User Facility supported under Contract DE-AC05-00OR22725. 
This research also used resources of the National Energy Research Scientific Computing Center,
a DOE Office of Science User Facility supported by the Office of Science of the U.S.
Department of Energy under Contract No. DE-AC02-05CH11231.

\balance


\bibliography{references}
\bibliographystyle{unsrt}  

\end{document}